\documentclass[conference]{IEEEtran}
\IEEEoverridecommandlockouts
\usepackage{cite}
\usepackage{amsmath,amssymb,amsfonts}
\usepackage{algorithm}
\usepackage{algpseudocode}
\usepackage{graphicx}
\usepackage{textcomp}
\usepackage[capitalise]{cleveref}
\usepackage{booktabs}
\usepackage{xcolor}
\usepackage{stfloats}

\def\BibTeX{{\rm B\kern-.05em{\sc i\kern-.025em b}\kern-.08em
    T\kern-.1667em\lower.7ex\hbox{E}\kern-.125em}}
    
\makeatletter
\newcommand{\linebreakand}{%
  \end{@IEEEauthorhalign}
  \hfill\mbox{}\par
  \mbox{}\hfill\begin{@IEEEauthorhalign}
}
\makeatother

\begin{document}

\title{Device-Free Human State Estimation using UWB Multi-Static Radios
\thanks{This work was done at the Samsung AI Center, Montreal.}
\thanks{\textsuperscript{*} Equal contribution.}
}

\author{
\IEEEauthorblockN{Saria Al Laham\textsuperscript{*}}
\and
\IEEEauthorblockN{Bobak H. Baghi\textsuperscript{*}}
\and
\IEEEauthorblockN{Pierre-Yves Lajoie\textsuperscript{*}}
\linebreakand
\IEEEauthorblockN{Amal Feriani}
\and
\IEEEauthorblockN{Sachini Herath}
\and
\IEEEauthorblockN{Steve Liu}
\and
\IEEEauthorblockN{Gregory Dudek}
}

\maketitle

\begin{figure*}[b]
    \centering
    \includegraphics[width=\textwidth,trim={0cm 10cm 0cm 0cm},clip]{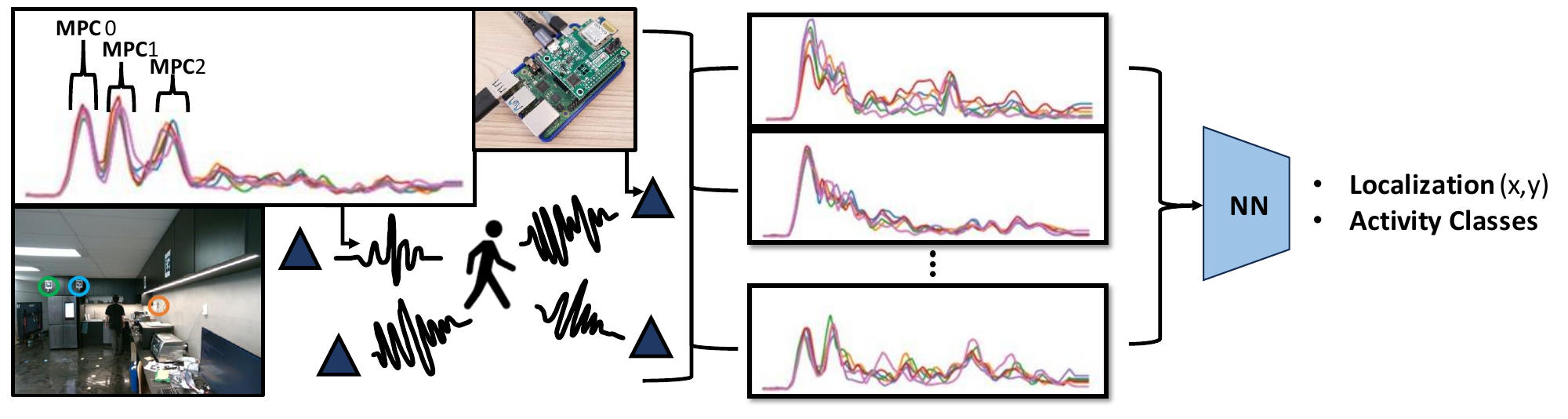}
    \caption{Overview of our device-free human-state estimation technique. UWB CIRs are accumulated and used to infer human localization and activity.}
    \label{fig:overview}
\end{figure*}

\begin{abstract}
 We present a human state estimation framework that allows us to estimate the location, and even the activities, of people in an indoor environment without the requirement that they carry a specific devices with them.  To achieve this ``device free'' localization we use a small number of
 low-cost Ultra-Wide Band (UWB) sensors distributed across the environment of interest. To achieve high quality 
 estimation from the UWB signals merely reflected of people in the environment, we exploit a deep network that can learn to make inferences.
 The hardware setup consists of commercial off-the-shelf (COTS) single antenna UWB modules for sensing, paired with Raspberry PI units for computational processing and data transfer.
We make use of the channel impulse response (CIR) measurements from the UWB sensors to estimate the human state - comprised of location and activity - in a given area. Additionally, we can also estimate the number of humans that occupy this region of interest. 
In our approach, first, we pre-process the CIR data which involves meticulous aggregation of measurements and extraction of key statistics. Afterwards, we leverage a convolutional deep neural network to map the CIRs into precise location estimates with sub-30 cm accuracy. Similarly, we achieve accurate human activity recognition and occupancy counting results. 
We show that we can quickly fine-tune our model for new out-of-distribution users, a process that requires only a few minutes of data and a few epochs of training.
Our results show that UWB is a promising solution for adaptable smart-home localization and activity recognition problems.
\end{abstract}

\section{Introduction}
We show how ultra-wide band sensors deployed across an environment can be used to infer the locations and activities of people moving around.  We do this by training a light-weight inference system that can correctly interpret the
complex UWB signals that arise from reflections in the environment.

Localization, as evidenced in the widely-used global positioning system (GPS), has had a titanic impact on many industries as well as on our daily lives. Yet, in indoor settings, localization technologies still have major limitations. To benefit from these technologies, users must generally wear or carry a device, such as a UWB tag or smartphone~\cite{lajoie2023peoplex}. In many circumstances,
however, continually carrying such devices is neither desirable nor practical.
For example, the requirement to wear a device at all times might be impractical or uncomfortable for patients in various health facilities, resulting in low compliance. Similarly, it is not ideal to ask inhabitants of a smart home to forego their comfort by a restrictive requirement to wear a specific device.
Furthermore, device-specific solutions require software compatibility between the deployed sensors in the environment and the user's devices. This becomes a major hurdle when deploying such technologies in commercial settings where the consumers wear a wide variety of devices from different manufacturers.
This motivates \emph{device-free} methods that enable the deployment of various services (e.g., localization, activity recognition) \emph{without} relying on users' mobile devices.

Currently, the best results 
for device-free localization of human 
subjects are  obtained using LIDAR or 
vision sensors and algorithms. Vision-based solutions, however, have several drawbacks that render them unacceptable in certain situations. Chiefly among these are privacy concerns that often preclude a continuously recording surveillance system in a home. Additionally, vision-based solutions, specially the deep-learning based ones, require expensive computational resources~\cite{Wang2023survey}. Moreover, their limited field of view and the occlusion of environmental objects often necessitates numerous cameras to obtain full coverage, further exacerbating the costs. 

To address these shortcomings, we opt to use Ultra-Wide Band (UWB) radios. UWB is a radio frequency (RF) technology that has recently attracted research attention in the robotics and localization communities~\cite{zafariSurveyIndoorLocalization2019}. As its name implies, UWB operates on a wide frequency band which allows for its transmitted pulses to be resolved very precisely in time. In typical localization usage, this fine time resolution is leveraged to accurately determine the signal time-of-flight (ToF) between UWB devices, and therefore their distances. However, in this work, we focus on device-free human state estimation by leveraging the fact that UWB signals, when travelling from one fixed transceiver to another, can encode information about the users when propagating through the environment.
Therefore, we can infer the state of the users by learning models that can extract it from the received signals.

\section{Background and Related Works}

\subsection{Indoor Localization}
The problem of locating a person, or a machine, indoors is one of the classic challenges of robotics and
machine vision, and has received attention from a vast number of researchers.  The most well-established version of the problem is known as simultaneous mapping and localization, where the objective is to both map the environment (often to establish the positions of fixed landmarks that may be semantic~\cite{li2019semantic} or geometric~\cite{stachniss2016simultaneous}), and then use such a map, or some set of landmarks, to estimate positions.  In this context the problem presupposes a device of interest as able to ``self-localize'' via such a process, but much of the same reasoning can be applied to the problem we consider here where a person must be localized with respect to the environment.  In the existing literature there have also been many methods that integrate information over time to achieve an unambiguous estimate~\cite{stachniss2016simultaneous,marinakis2006probabilistic,rao2007randomized}.
One important sub-theme is the use of ambient WiFi signals to infer people's positions, based on how these signals are reflected.  Approaches to this problem began with
geometric inference and
signature matching, and were often based on deep learning and/or auto-correlation~\cite{ladd2002robotics,siddiqi2003experiments,xu2020pressense,chen2022fidora,chen2020fido}.

\subsection{Ultra-Wide Band}
Ultra-Wide Band (UWB) is a wireless communication protocol \cite{ieee_uwb} that uses a broad spectrum of frequencies, typically exceeding 500 MHz. The wide bandwidth gives the UWBs unique advantages, particularly in terms of localization and short-range data transfer. UWB is renowned for its high precision in indoor localization (e.g., $30\;$cm and less). This precision comes from operating on a wide range of frequencies, which reduces the multipath fading effects where signals bounce off surfaces and objects. The multipath components (MPCs) implicitly carry information about the behavior of the surrounding environment. Each MPC arrives at the receiver side at different times due to bouncing off walls, floors and other objects, a phenomenon known as multipath propagation. These MPCs construct the channel impulse response (CIR), which is a complex discrete signal given by:
\begin{equation}
h(\tau) = \sum^{K}_{i=0}\alpha_i\cdot\delta(\tau-\tau^d_i);
\label{eq:channel_model}
\end{equation}
where $K$ is the number of paths, $\alpha_i \in \mathbb{C}$ and $\tau^d_i$ are the gain and the time delay of the $i$-th path, respectively. 
The first MPC, $\alpha_0$, represents the line-of-sight (LoS) path if it exists. The CIR can be estimated via algorithms such as deconvolution \cite{deconv_muqaibel2002uwb} when two UWB sensors communicate a preamble sequence that is known in advance. The power of the CIR lies in its ability to capture the RF fingerprint of the environment through the MPCs and transform it into rich data that is ripe for analysis, whether for localization or activity recognition.

\subsection{Device-Free Localization}
Device-free, or passive, localization aims to position and track human users in an indoor environment without requiring them to carry any electronic device~\cite{Rathin2019taxonomy}. Relying only on ambient signal analysis, it has emerged as an alternative to conventional localization methods. Techniques like signal strength, and ToF are common, though challenges such as signal interference and environmental factors affect accuracy.
Device-free localization also include a range of non-RF techniques~\cite{Fakhrul2021nonRF} from infrared sensing to human-object interaction detection~\cite{ruan2016humanobject}.

Among the most popular approaches, WiFi-based device-free localization solutions use existing network infrastructure, ubiquitous in indoor environments, to locate subjects~\cite{LiHedley2019passive}. Thus, those approach are cost-effective and easy to deploy but faces issues like interference and limited precision. 
In particular, given that learning-based techniques are currently the state-of-the-art, most approaches suffer from generalization on new users with different body shapes or daily changes of the environment, such as opening/closing doors. Thus, there is a need for fine-tuning, or domain adaptation, to obtain robust results consistently as shown in Chen~et~al.~\cite{chen2020fido,chen2022fidora}.

Thanks to its larger bandwidth and precise time resolution, UWB is increasingly considered as an alternative to WiFi for higher accuracy solutions~\cite{wang2017}. The improved performance justifies the cost of deploying specialized hardware.
UWB-based approaches usually detect time variations on the measured CIRs between transmitters and receivers. As in Jovanoska~et~al.~\cite{Jovanoska2013}, the time variations, caused by humans in the environment, are compared with previously recorded background data (i.e., empty room) to extract range information which can be fused using maximum likelihood estimation. In the same vein, Lei~et~al.~\cite{lei2021} processed UWB ToF measurements through clustering and geometric filtering to obtain robust multi-target localization estimates. Interestingly, novel CIR-based approaches have the opportunity to leverage information beyond the first transmission path, used for ranging, and extract meaningful data from subsequent components. While Cimdins~et~al.~\cite{Cimdins2020} extracted features from the signal's multipath components, we learn the relevant features using a neural network.
Probably the closest technique to our work is Li~et~al.~\cite{LiMultistatic2022} where the authors presented a multi-static UWB radar setup leveraging mean and variance information from accumulated CIRs to learn a CNN-based model for localization.

\subsection{Device-Free Activity Recognition}

Human activity recognition (HAR) seeks to recognize and classify human activity. HAR exists in a wide gamut of defined activities and sensing modalities. The granularity of what defines human activity ranges from high-level motion information such as standing or walking \cite{sharmaDeviceFreeActivityRecognition2019, zhengliangDatasetHumanMotion2021a} to highly granular activities of daily living (ADL) \cite{bouchardActivityRecognitionSmart2020,lafontaineDenoisingUWBRadar2023}. For a comprehensive review of device-free HAR, we refer the reader to the survey by Yang et. al~\cite{yangDeepLearningTransfer2022}.

While past work has shown the promise of wearable UWB tags in human activity detection~\cite{chengActivityRecognitionLocalization2020}, we focus on device-free UWB-based activity detection where participants do not wear any device. 
Instead, device-free methods rely on reflected UWB signals, which are sensitive enough to accumulate subtle information about the environment and its inhabitants. The central challenge, then, is to extract this information from noisy UWB CIRs to infer human activity.

Zhengliang et. al~\cite{zhengliangDatasetHumanMotion2021a} proposed a UWB radar motion detection dataset that classifies human activity into standing, walking, and `no humans'. They showed effective classification using a CNN-based architecture.
Sharma et al.~\cite{sharmaDeviceFreeActivityRecognition2019} compared the HAR performance of UWB radios to WiFi CSI. Their findings showcased that a variety of ensemble and deep learning techniques perform better when using UWB CIR when compared to WiFi CSI.
Bouchard et al.~\cite{bouchardActivityRecognitionSmart2020} collected ADLs in a controlled smart home environment using a UWB radar. For HAR, they used K-nearest neighbours (KNN), classification and regression tree (CART)~\cite{breiman1984classification}, AdaBoost \cite{ADABOOST}, and random forest techniques. In follow up work, LaFontaine et al.~\cite{lafontaineDenoisingUWBRadar2023} show that denoising CNN autoencoders (AE) can be applied to preprocess the UWB signal and significantly boost the classification performance of a DNN-based method. 
Nouri et al.~\cite{nooriUltraWidebandRadarBasedActivity2021} use principal component anaylysis (PCA) and linear discriminant analysis (LDA) to obtain features from UWB signals which are then injested by an LSTM~\cite{hochreiter1997long} architecture for activity recognition on a 5-class activity dataset.


\section{Data Collection}
The following section describes our data collection pipeline, which is a crucial component for the success of our approach and could inspire future implementations.

\begin{figure}
    \centering
    \includegraphics[width=\columnwidth]{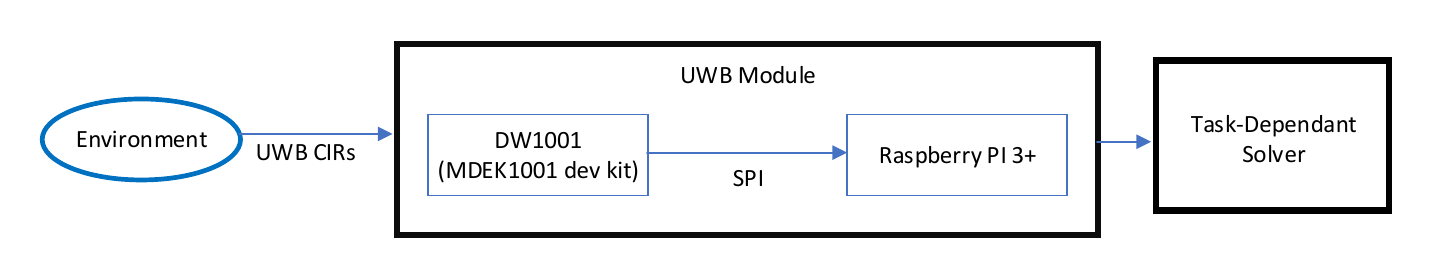}
    \caption{Diagram of UWB module developed at SAIC-Montreal}
    \label{fig:uwb_module}
\end{figure}

\subsection{Hardware Setup}
To address the need for real-time efficient UWB CIR sensing for our deep learning pipeline, we design a custom UWB module, detailed in~\cref{fig:overview,fig:uwb_module} to collect CIRs from the UWB chip and transmit this data to a central processing hub. The module consists of a UWB chip, an antenna, and a network-capable compute unit. In our design, we use the DW1001~\cite{DW1000} chip with an onboard antenna, which is a commonly used UWB platform for research purposes~\cite{LiMultistatic2022} as it exposes the internal CIR buffers to the user.

To allow rapid network data transfer, we attach this unit to a Raspberry PI 4+~\cite{raspberrypi} (RPI). In the setup, each UWB module acts as a transmitter (Tx) and a receiver (Rx), where it frequently broadcasts beacon packets to other UWB modules, and wait for their broadcasts. We employ a random access protocol, specifically, a simplified ALOHA~\cite{ALOHA1970}, where each UWB module send a beacon packet and then wait for a random period of time (e.g., between 0 and 100 $\mu s$) or until it receives a beacon packet before sending another.
When decoding the packet, the DW1001 chip can extract CIR from the preamble sequence using a proprietary Leading Edge Detection (LDE) algorithm \cite{s20061599}. It exposes a 1016 samples long CIR buffer, with approximately 1~$ns$ delay between each sample for each valid preamble reception. 
However, only the CIR values surrounding the first path (FP) index, i.e., the first MPC, are of our interest. 
Therefore, we choose to report 58 CIR samples for a single CIR measurement, which are sufficient to capture the 
RF fingerprint of the environment up to 15 meters, starting from 3 samples before the first MPC. Each CIR sample is a complex number, and represented by four bytes, two bytes for the real part, and the other two bytes for the imaginary part. As such, each CIR measurements is 232 bytes. Additionally, we report metadata that consist of: beacon sequence number, FP index, fractional part of the FP, preamble count, Tx ID, and Rx ID, 2 bytes each, in addition to 6 bytes for the internal clock timestamp, resulting in 18 bytes in total for the metadata, and a total of 250 bytes to transfer to the RPI. The used UWB settings for the DW1001 chip is detailed in Table \ref{tab:uwb_specs}.

\begin{table}[thbp]
\centering
\caption{UWB settings of the DW1000 chip}
\begin{tabular}{ll}
\hline
Parameter                        & Value      \\ \hline
IEEE 802.15.4.a channel Number   & 5          \\
Carrier frequency                & 6489.6 MHz \\
Bandwidth                        & 499.2 MHz  \\
Pulse repetition frequency       & 64         \\
Preamble length                  & 256        \\
Preamble acquisition chunk size & 16         \\
Tx preamble code sequence        & 10         \\
Physical layer header mode       & Standard   \\ \hline
\end{tabular}
\label{tab:uwb_specs}
\end{table}

To obtain high UWB sensing rates, we must minimize the delay between reception and transmission which is dominated by the data transfer time of the CIR to the RPI. We found that using the fastest data bus on the chip, the serial peripheral interface (SPI) allowed us to achieve approximately 8ms per UWB beacon broadcast.

The UWB modules are networked in a decentralized manner to a desktop computer which acts as a central processing hub. We achieve this by leveraging the Robot Operating System 2 (ROS~2)~\cite{macenski2022ROS2}. ROS~2 is a decentralized publisher-subscriber networking and compute framework which was developed by the robotics community to address their heterogeneous sensing and compute needs. In particular, we leverage ROS~2 ability to asynchronously and rapidly transfer data to the centralized compute server for real-time computing and storage~\cite{bedardROS2_2023}, which is critical in distributed sensor networks~\cite{lajoie2022ROS2}.
\begin{figure}[thbp]
    \centering
    \includegraphics[width=\columnwidth]{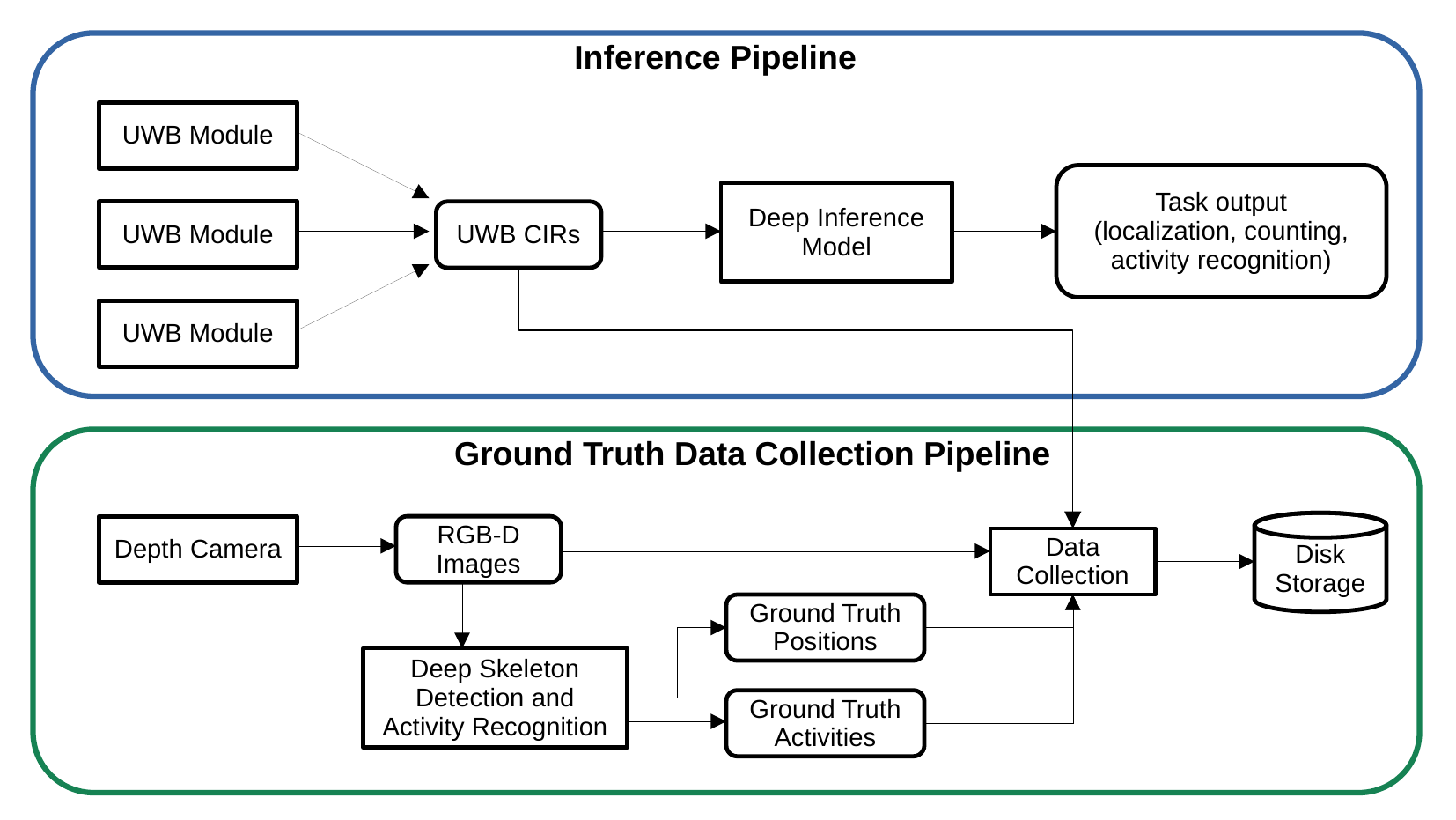}
    \caption{ROS~2-based sensor network pipeline. Circles represent “nodes” which collect data or perform computation tasks. Rectangles represent "topics" which are live information stores for nodes to write to and read from to perform their tasks.}
    \label{fig:ROS2_pipeline}
\end{figure}
ROS~2 manages its publisher-subscriber communication model through the concept of `topics' and `nodes'. In ROS~2, topics are repositories of information such as sensor data. This data can be read or written to via ROS~2 nodes, which can optionally both subscribe and/or publish to said topics. In our implementation, each UWB module runs a sensing node which publishes the collected CIR data to a topic, which can be retrieved later by the deep learning model node for training or inference purposes. A high-level overview of the inference pipeline is summarized in~\cref{fig:ROS2_pipeline}.

\begin{table}[t]
    \vspace{4mm}
    \centering
    \caption{Performance details for best-trained model for the purpose of indoor device free human localization. Metrics in meters. 
    }
    \begin{tabular}{@{}lc@{}}\toprule
Mean Localization Error & 0.209\\
Standard deviation & 0.165\\
Median Localization Error  & 0.173\\
80th Percentile error & 0.334\\
Maximum error  & 1.092
    \\\bottomrule
    \end{tabular}
    \label{tab:loc_results}
    \vspace{-5mm}
\end{table}

\begin{figure}[thbp]
    \centering
    \includegraphics[width=\columnwidth]{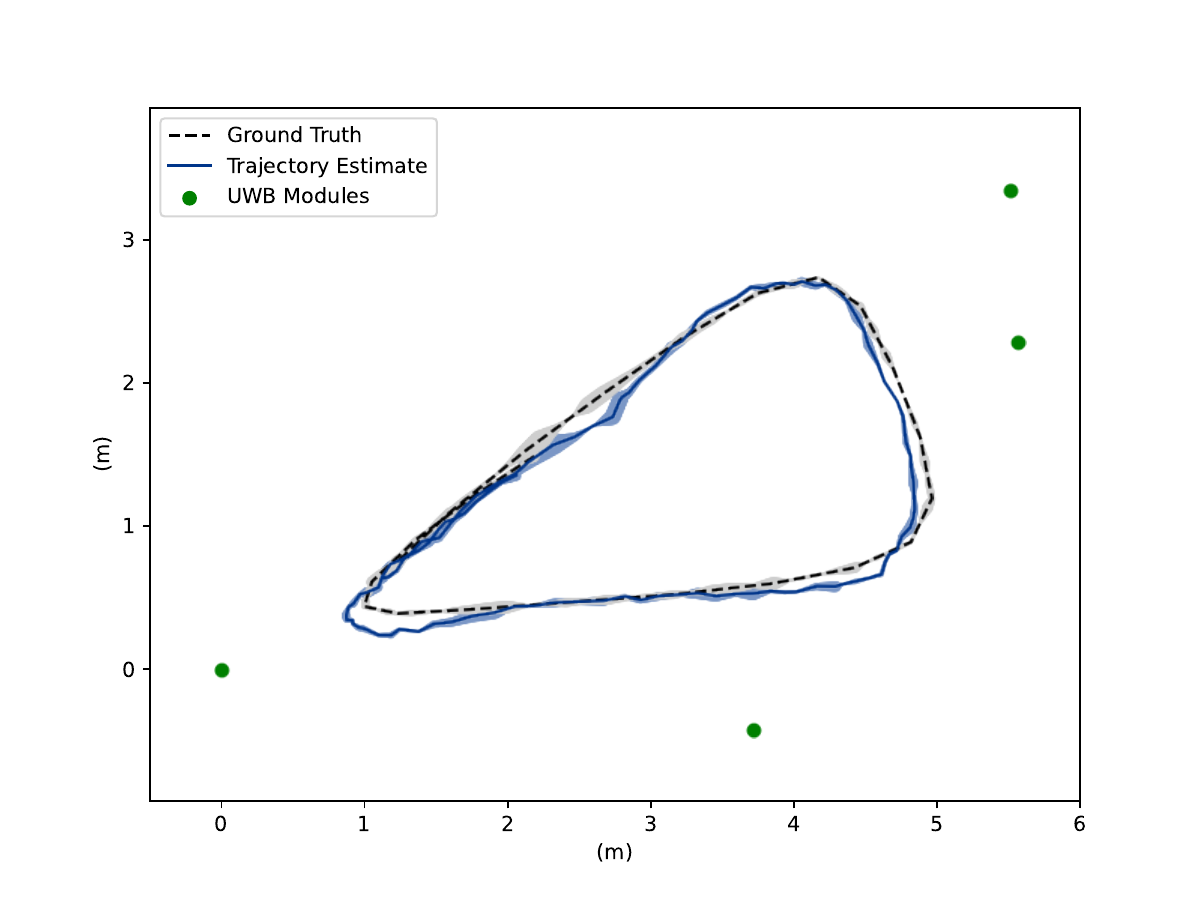}
    \caption{Sample of the trajectory output from our device-free localization model. The ground truth trajectory is in black while the estimated trajectory (temporally smoothed) is in blue. The green dots show the real-world locations of the UWB sensing modules.}
    \label{fig:loc-results-viz}
\end{figure}

\begin{figure}[thbp]
    \centering
    \includegraphics[width=\columnwidth,trim={5cm 3cm 3cm 2cm},clip]{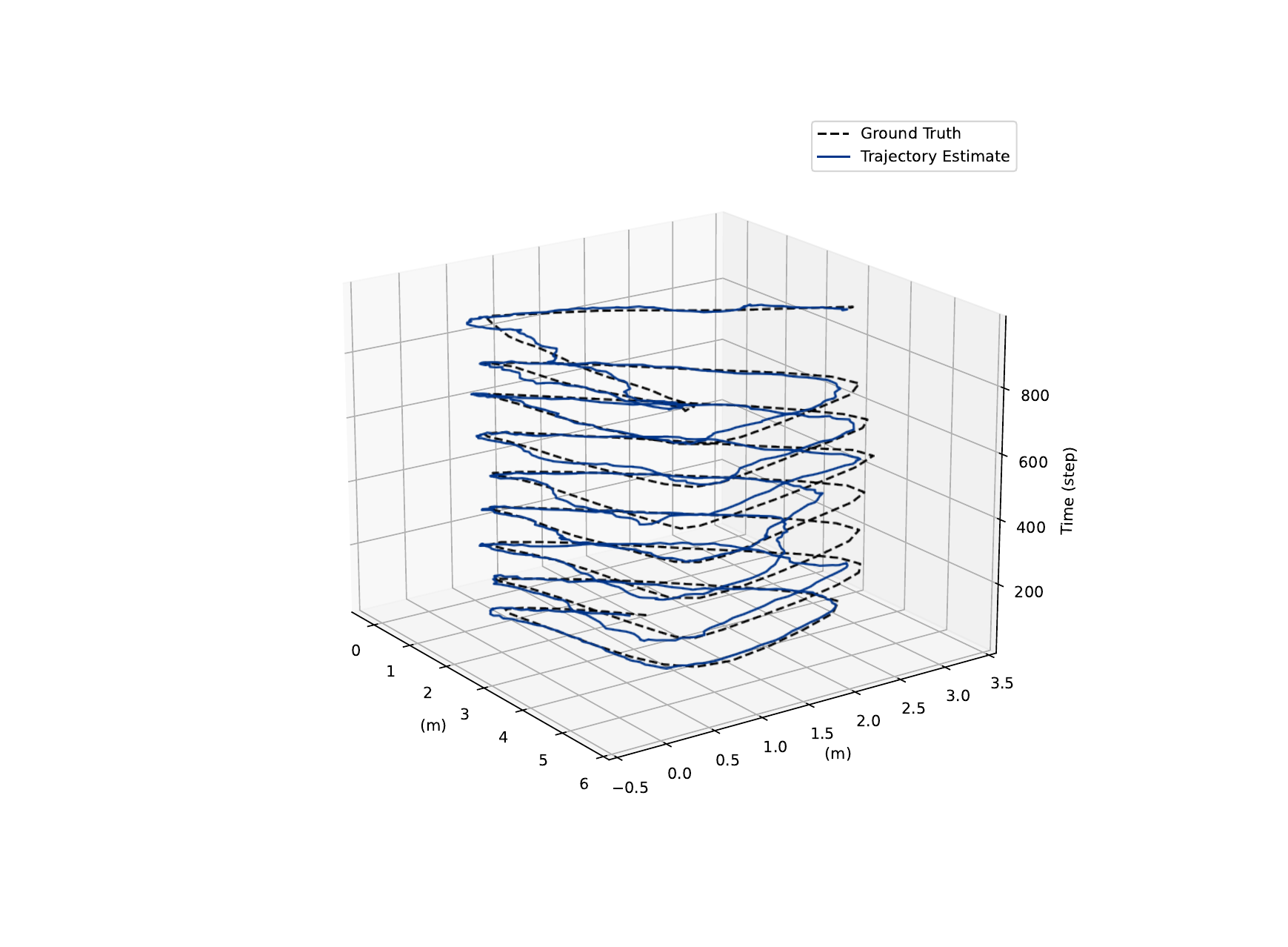}
    \caption{Sample of the trajectory output from our device-free localization model. The ground truth trajectory is in black while the estimated trajectory (temporally smoothed) is in blue. Time on Z-axis.}
    \label{fig:loc-results-viz3d}
\end{figure}

\begin{figure}[thbp]
    \centering
    \includegraphics[width=\columnwidth]{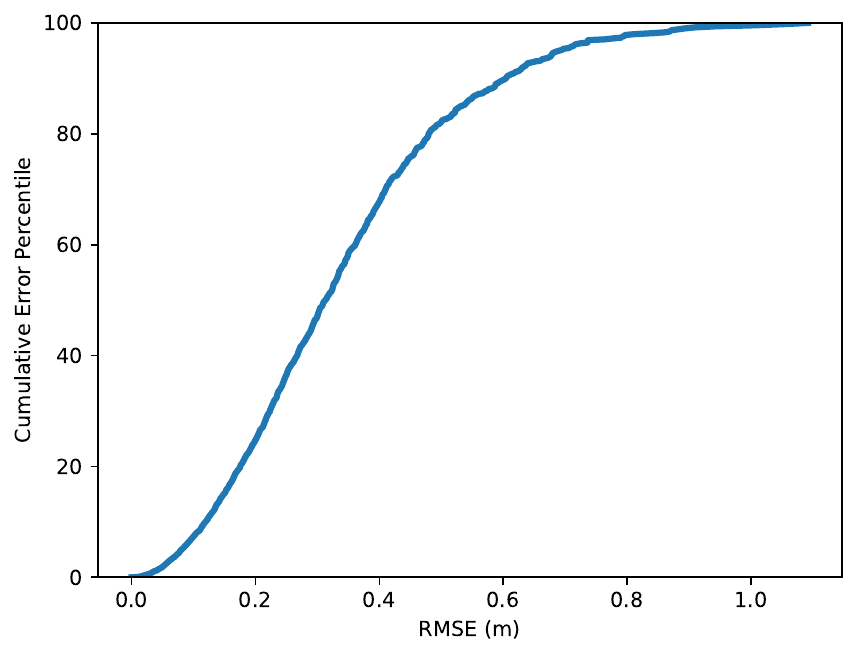}
    \caption{The CDF plot of localization errors in our real world experiments.}
    \label{fig:error-cdf}
\end{figure}

\subsection{Ground Truth}

To train our model, we require the ground truth location and human activity labels which we obtain through a camera-based setup. Note that the camera is only used for ground truth collection and does not play a role in the inference portion. To obtain precise human positions, we extract human skeletons from an RGBD camera using the Pyskl library~\cite{duan2022pyskl}. When properly calibrated, the depth information allows us to reproject the detected skeleton keypoints back into 3D space. Then we define the location of a person as the average of their 3D reprojected keypoint positions. 

A key problem in ground truth collection is the time synchronization of all involved elements, including UWB modules and the camera. To do this, we run a local network time protocol (NTP) server to time synchronize all our devices' clocks. Furthermore, the ground truth collection instruments are all implemented in ROS~2, unifying the inference and ground truth pipelines and simplifying the synchronization problem. Our unified inference and ground truth pipeline is summarized in Fig.~\ref{fig:ROS2_pipeline}.

The ground truth data for activity recognition and occupancy counting is manually labelled, as is discussed in Subsection~\ref{sec:experiment_design}.

\subsection{Experiment Design}
\label{sec:experiment_design}

We implement a straightforward experimental design for both ground truth and testing data collection. For the localization task, we introduce the participant to the collection area. Participants are then instructed to walk in the area in a free-form manner without any restriction on the nature of their chosen path nor their speed. In practice, we found that the participants expressed enough variety in their motions (including periods of standing still) and produced enough coverage of the area that no further instructions were required. 
For the activity recognition task, participants are asked to perform a certain activity for the duration of the collection period, such as walking or sitting. Note that for stationary activities, we vary the location so as to collect location-independent data for training and testing. 
A similar routine is prescribed for occupancy counting, where the participant(s) are instructed to be present in the collection area without any specific indication as to their motion or activity for the duration of data collection.

\section{Methodology}

\subsection{Data Preprocessing}
Recall that in our experiment, each UWB module frequently receives and transmits a beacon packet to other UWB modules. When decoding the packet, the UWB node can extract CIR measurements. We employ 4 UWB modules, resulting in 12 communications links in total. A single CIR measurement in a single link consists of 58 samples with approximately 1 $ns$ resolution. The CIR signal envelope should be constant in a stable environment. However, there exist software/hardware impurities in the used DW1001 chip, specifically the proprietary LDE algorithm,  which results in inaccurate CIR measurements. As such, we need to aggregate multiple CIR measurements for each link to produce a stable one. 
Therein, for each link, we obtain M consecutive CIR measurements that go into a preprocessing block. In the preprocessing, first, we align the CIR measurements by making use of the fractional FP part that was reported in the metadata, which represents the offset from the FP index. Each single CIR measurement have different offset. The offset can be one of 64 value in the range of $[0,~64]$ with $\frac{1}{64}$ $ns$ step. The alignment process and algorithm is detailed in \cite{s20061599, 7805795}. After the alignment of the M CIR measurements, we calculate two signals, namely, the running mean $h_m(\tau)$ and variance $h_v(\tau)$ of the collected M CIR measurements. Such signals are important to understand and visualize the deviations and changes within the CIR measurements over time. The mean and the variance are then upsampled to 500 samples using linear interpolation for better visualization and finer granularity. The mean and variance calculation and upsampling are then repeated for every other consecutive M CIR measurements. For more details about the preprocessing of the CIR measurements, one can refer to \cite{s20061599, LiMultistatic2022}. 
Therein, we stack the means and the variances of the CIR measurements within a time window $\mathcal{T}$ (e.g., 1 second), and the number of means and variances that we get in the time window are denoted by $C$, i.e., we have $C$ means and variances per $\mathcal{T}$. Afterwards, the data from all the links is stacked together, resulting in a single CIR data point with shape $(12, 2, C, 500)$, where 12 is the number of communication links, 2 for mean and variance, $C$ is the number of means and variances calculations in time window of $\mathcal{T}$, 500 is the number of samples in the mean and variance. The labeling process is as follows, for each data point, we take the true position with a time stamp that is the closest to the time stamp at the end of the time window $\mathcal{T}$

\subsection{Deep Learning approach}
Our approach is to learn a many-to-one mapping, that is, we want to learn the mapping of a sequence of CIR measurements from all the links to a single position estimate $\hat{p} $, i.e.,:
\begin{equation}
\hat{p} = f(\bigcup_{i\in \mathcal{I}}\bigcup_{t \in \mathcal{T}} \left [  h^{i, t}_m, h^{i, t}_v (\tau) \right ])
\end{equation}
where $h^{t,i}_m (\tau)$ and $h^{t,i}_v (\tau)$ are the CIR measurements mean and variance for link $i$ at time $t$, respectively, $\mathcal{I}$ is the set of UWB communication links, $\mathcal{T}$ is the CIR measurements collection time window, and $f(.)$ is the mapping function.
\begin{figure}[tbhp]
    \centering
    \includegraphics[width=\columnwidth]{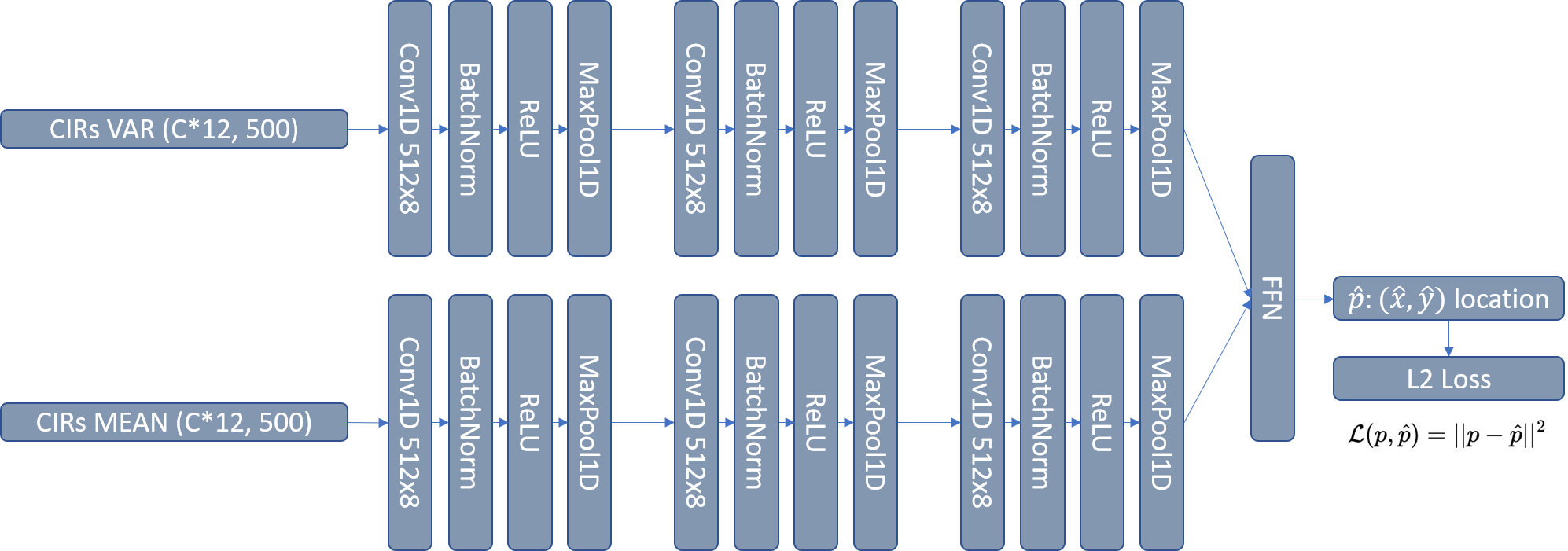}
    \caption{Network Architecture}
    \label{fig:cnn2tails}
\end{figure}
We believe this is one of the first end-to-end approach of its kind, where we input a sequence of preprocessed CIRs directly into a Deep Learning (DL) model, and expect a location estimate. We rely on the fact that the model can learn the geo-temporal features from the CIRs without any external feature extraction. 
Our DL model consist of two streams of convolutional layers, one for the CIRs variance and the other for the CIRs mean signals. As described in~\cref{fig:cnn2tails}, each stream consists of multiple convolutional, batch normalization, and max pooling layers, and the output of the two streams is then concatenated and fed into a fully connected Feed Forward Network (FFN), which outputs a position estimate. We adopt the minimization of the L2 loss as our objective, given by:
\begin{equation}
\mathcal{L}(p, \hat{p}) = ||p-\hat{p}||_2 
\end{equation}
For other tasks such as people counting and activity recognition, the output layer of the FFN works as a classifier, where the output is one class (e.g., standing, walking, etc.), and the objective is substituted with the Cross Entropy loss.

\section{Results}

\subsection{Localization}
Our experimental localization results are presented in Table~\ref{tab:loc_results}. With a mean localization accuracy of 0.213 m, our method can achieve precise indoor localization. Furthermore, the method is robust even up to the 80\textsuperscript{th} percentile, with only a 0.06 m increase in error. However, it should be noted that some outliers cause large errors. We hypothesize that these errors are due to environmental factors such as non-line-of-sight propagation or material interactions with the reflected UWB signals. The largest localization error, in such scenarios, was recorded to be 1.21m.

\begin{table}[t]
    \vspace{4mm}
    \centering
    \caption{Finetuning device free localization model on perturbed environment and unseen human participants.}
    \begin{tabular}{@{}lcccccc@{}}\toprule
Test sample & \# samples  & Mean  & STD  & Median  & 80th Perc. & Max \\
Person1  & 285  & 0.410  & 0.290  & 0.332  & 0.630  & 1.39 \\
Person2  & 403  & 0.345  & 0.198  & 0.311  & 0.505  & 1.11 \\
Person3-1   & 355  & 0.400  & 0.240  & 0.359  & 0.579  & 1.30 \\
Person3-2  & 393  & 0.351  & 0.239  & 0.290  & 0.524  & 1.49  \\
Person4   & 1213  & 0.314  & 0.243  & 0.254  & 0.481  & 1.26
    \\\bottomrule
    \end{tabular}
    \label{tab:finetuning_counting_results}
\end{table}

\begin{figure}[tbhp]
    \centering
    \includegraphics[width=\columnwidth]{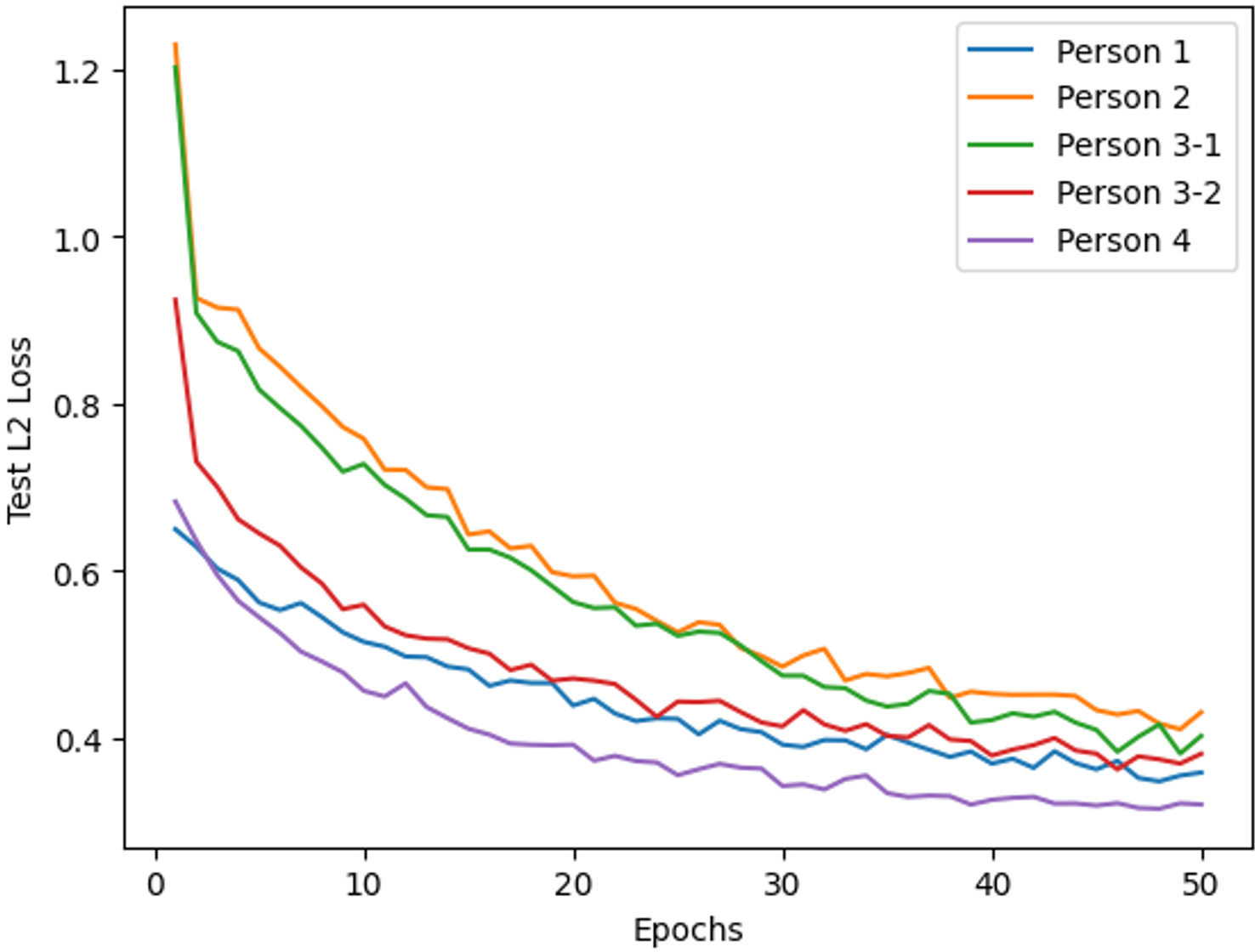}
    \caption{Finetuning progress. Ten epochs translate to around one minute of training time. Model performance can be restored in about five minutes of training, even when using never seen human participants.}
    \label{fig:people-finetuning}
\end{figure}

For a qualitative evaluation of our localization, we show the comparison between ground truth location and inferred positions in Fig.~\ref{fig:loc-results-viz}. While the inferred trajectory can be seen to closely follow the ground truth for the most part, there exist locations that cause significant error (see right side of Fig.~\ref{fig:loc-results-viz3d}). It is the apparent location dependency of these errors that leads us to our earlier error hypothesis based on environmental factors.

\begin{table}[t]
    \vspace{4mm}
    \centering
    \caption{Occupancy counting testing results.}
    \begin{tabular}{@{}lccc@{}}\toprule
Count & Precision & Recall & F1-score\\
1  & 0.99 & 0.98 & 0.99\\
2  & 0.99 & 1.00 & 0.99\\
3 & 0.99 & 0.99 & 0.99\\
4  & 1.00 & 0.97 & 0.98\\
Macro-avg  & 0.99 & 0.98 & 0.99 \\
    \midrule
Accuracy && 99\% &
    \\\bottomrule
    \end{tabular}
    \label{tab:occupancy_counting_results}
\end{table}

Since UWB CIRs are sensitive even to the smallest environmental changes, it is difficult to train a deep learning model for complete generalization. In fact, the accumulation of environmental alterations which result from daily usage is enough to significantly reduce the localization accuracy after a few days. Similarly, the model has difficulty generalizing to out-of-distribution individuals, as well as in-distribution individuals with sufficiently different state (e.g. clothing or subtle physiological changes from a few days).

A straightforward method to combat this deficiency is to simply tune the model with fresh data. This begs the question: how much data is required for this training, and how much training is required? We answer this question by performing experiments where a stale model is fine-tuned using only five minutes of data on four novel individuals. The results of this experiment are visualized in Fig.~\ref{fig:people-finetuning}. It is apparent that our model can restore its performance using a few minutes of data in as much training time. In the absence of other generalization strategies, this could be a valid strategy to mitigate model staleness.

\begin{table}[t]
    \vspace{4mm}
    \centering
    \caption{Device free activity recognition results.}
    \begin{tabular}{@{}lccc@{}}\toprule
Activity  & Precision & Recall & F1-score   \\
Moving  & 1.00 & 1.00 & 1.00 \\
Standing  & 0.98 & 0.91 & 0.94 \\
Sitting  & 0.89 & 0.97 & 0.93 \\
Macro-avg  & 0.96 & 0.96 & 0.96  \\
    \midrule
Accuracy && 97\% &
    \\\bottomrule
    \end{tabular}
    \label{tab:device_free_har_results}
\end{table}

\subsection{Occupancy Counting and Activity Recognition}
The occupancy counting results are presented in Table \ref{tab:occupancy_counting_results}. The proposed method was successful in counting up to 4 people in a single area. It achieved an average accuracy of 99\%, a precision of 0.99, a recall of 0.98, and an F1-score of 0.99. The occupancy counting problem is rather simple compared with the other problems since it mainly involved identifying different peaks in the CIR measurements that reflects the count.
\subsection{HAR results}
The HAR results are shown in Table \ref{tab:device_free_har_results}. As expected, the proposed algorithm can successfully identify whether the subject is moving, standing, and sitting, with an average accuracy of 97\%. Differentiating between activities such as moving and standing is a simple task using the spatio-temporal features that the CNN extracts from the CIR measurements. However, we believe that with more complex actions including different actions with smaller movements (i.e., brushing teeth), the  HAR problem becomes very challenging considering our minimally invasive setup. We leave this challenging task for future work.

\section{Conclusion}
In summary, our exploration into device-free UWB-based indoor localization provides insights on the design of effective indoor localization systems and the associated challenges. We believe our work and results will be helpful for both the research and industrial communities.

\bibliographystyle{IEEEtran}
\bibliography{main.bib}

\end{document}